\documentclass[preprint,superscriptaddress,amsmath,amssymb,aps,prab,longbibliography,titlepage]{revtex4-2}

\usepackage{graphicx}
\usepackage{natbib}
\usepackage{hyperref}
\usepackage{cleveref}
\usepackage{mathrsfs}
\usepackage{gensymb}
\usepackage{upgreek}
\usepackage[usenames,dvipsnames]{xcolor}
\usepackage{mathptmx}

\hypersetup{colorlinks=true,citecolor=blue,urlcolor=blue}

\newcommand{\Wcmsqd}{\mathrm{W}\text{cm}^{-2}}
\newcommand{\rmd}{\mathrm{d}}
\newcommand{\micron}{{\upmu\mathrm{m}}}

\newcommand{\abs}[1]{\left| #1 \right|}
\newcommand{\Ecrit}{E_\text{cr}}
\newcommand{\ncrit}{n_\text{cr}}

\DeclareMathOperator{\Ai}{Ai}
\renewcommand{\vec}{\mathbf}

\begin{document}

\title{Self-absorption of synchrotron radiation in a laser-irradiated plasma}

\author{T. G. Blackburn}
\email{tom.blackburn@physics.gu.se}
\affiliation{Department of Physics, University of Gothenburg, SE-41296 Gothenburg, Sweden}
\author{A. J. MacLeod}
\author{A. Ilderton}
\author{B. King}
\author{S. Tang}
\affiliation{Centre for Mathematical Sciences, University of Plymouth, Plymouth, PL4 8AA, United Kingdom}
\author{M. Marklund}
\affiliation{Department of Physics, University of Gothenburg, SE-41296 Gothenburg, Sweden}

\date{\today}

\begin{abstract}
Electrons at the surface of a plasma that is irradiated by a laser with intensity
in excess of $10^{23}~\Wcmsqd$ are accelerated so strongly that they emit bursts of
synchrotron radiation.
Although the combination of high photon and electron density and
electromagnetic field strength at the plasma surface makes particle-particle interactions possible, these interactions are usually neglected in simulations of the high-intensity regime.
Here we demonstrate an implementation of two such processes: photon absorption and stimulated emission.
We show that, for plasmas that are opaque to the laser light, photon absorption would cause complete depletion of the multi-keV region of the synchrotron photon spectrum, unless compensated by stimulated emission.
Our results motivate further study of the density dependence of QED phenomena in strong electromagnetic fields.
\end{abstract}

\maketitle

\section{Introduction}

Radiation emission from accelerated electrons is a ubiquitous feature of regions of strong electromagnetic field.
In astrophysical environments~\cite{harding.rpp.2006}, or in laser-matter interactions at the high-intensity frontier~\cite{danson.hplse.2019}, the fields can be so strong that the interactions must be described within the framework of quantum electrodynamics (QED)~\cite{erber.rmp.1966,marklund.rmp.2006,dipiazza.rmp.2012}.
Experiments at the next generation of high-intensity laser facilities~\cite{papadopoulos.hpl.2016,weber.mre.2017,gales.rpp.2018} will produce high-energy $\gamma$ rays via quantum synchrotron emission (also called nonlinear Compton scattering) in a variety of geometries~\cite{waltz.pf.1978,ridgers.prl.2012,li.pop.2014,nerush.ppcf.2015,stark.prl.2016}.
Particle-in-cell (PIC) simulations, extended to include the one-particle to two-particle (`1 to 2') strong-field QED processes of photon emission and electron-positron pair creation~\cite{ridgers.jcp.2014,gonoskov.pre.2015}, play an essential role in modelling these interactions.
However, for every emission process, there is a corresponding absorption process.
To date, the inverse (`2 to 1') processes of one-photon absorption~\cite{ilderton.prd.2019} and pair annihilation to one photon~\cite{voroshilo.lp.2010,tang.pra.2019} have been neglected in PIC simulations.

Here we consider the effect of one-photon absorption in a scenario where the photons are absorbed by the same population of relativistic electrons that emitted them.
In an astrophysical context, this phenomenon is known as \emph{synchrotron self-absorption}~\cite{longair}.
It leads to a steep cutoff at low frequency in the emission spectra from, e.g., supernovae~\cite{chevalier.apj.1998}, gamma-ray burst afterglows~\cite{granot.apj.1999,piran.rmp.2005}, and black hole X-ray binaries~\cite{fender.mnras.2019}.
In principle, the irradiation of a solid target by a laser of intensity $\gtrsim 10^{23}~\Wcmsqd$ is a platform for studying self-absorption, because of the combination of strong electromagnetic field, high electron density, and high photon density at the plasma surface.
A consistent treatment of photon absorption must include stimulated emission, which is the competing, induced process.
To do so, we construct a cross section for stimulated emission in QED that is valid within the locally constant, crossed fields approximation;
to the best of our knowledge, a cross section from QED has not previously been reported.
We present an implementation of both as binary interactions between macroparticles in a PIC code.
Simulating a laser-plasma interaction, we find that while photon absorption suppresses the multi-keV region of the synchrotron spectrum, this suppression is countered by stimulated emission.
Our results demonstrate that it is feasible to include strong-field particle-particle interactions in studies of laser-driven plasmas.

\section{Induced processes}

The following master equation determines the evolution of the number of photons, $N(\vec{k})$, with momentum $\vec{k}$~\cite{melrose}:
    \begin{equation}
    \frac{\rmd N(\vec{k})}{\rmd t} =
        \int\! \frac{\rmd^3\vec{p}}{(2\pi)^3}
            w(\vec{p},\vec{k})
            \left\{
                [1 + N(\vec{k})] f(\vec{p})
                - N(\vec{k}) f(\vec{p} - \vec{k})
            \right\}.
    \label{eq:MasterEquation}
    \end{equation}
Here $w(\vec{p},\vec{k})$ is the rate at which an electron with momentum $\vec{p}$ emits photons with momentum $\vec{k}$ and $f(\vec{p})$ is the electron distribution function, defined by $\rmd N_e = f(\vec{p}) \,\rmd^3\vec{p}/(2\pi)^3$.
(We use units such that $\hbar = c = 1$ throughout).
The first term in square brackets on the RHS of \cref{eq:MasterEquation} describes `spontaneous emission', which is the quantum synchrotron emission already included in laser-plasma simulations~\cite{ridgers.jcp.2014,gonoskov.pre.2015}.
The following two terms correspond, respectively, to the induced processes of stimulated emission and photon absorption.
Unlike spontaneous emission, they depend on the density of photons already present.
All three processes depend on the electron and photon momenta, $p^\mu$ and $k^\mu$, and the strength of the electromagnetic field $F_{\mu\nu}$, which is implicit in $w(\vec{p},\vec{k})$.
Stimulated emission does not represent an interaction in the same sense that absorption does: a photon is not absorbed and re-emitted, for example.
It is a consequence of the fact that photons are bosons, which means that the phase space for the emission of a photon with momentum $\vec{k}$ is enhanced by the presence of photons with the same momentum.
(For fermions, the equivalent phenomenon, Pauli blocking, would be de-enhancing.)

Conservation of momentum means that an electron in vacuum cannot absorb radiation without some associated emission of radiation. Absorption can occur, however, for an electron in a background electromagnetic field $F_{\mu\nu}$ (where the required emissions appear as `absorption' of negative frequency modes from the background~\cite{ilderton.prd.2019}).
If the field is weak compared to the critical field of QED, $\Ecrit = m^2 / e$~\cite{heisenberg.zp.1936,schwinger.pr.1951}, and if it varies sufficiently slowly such that quantum processes can be considered to be instantaneously constant, the interaction is controlled by the quantum parameters $\chi_e = \abs{F_{\mu\nu} p^\nu} / (m \Ecrit)$ and $\chi_\gamma = \abs{F_{\mu\nu} k^\nu} / (m \Ecrit)$, where $p$ and $k$ are the electron and photon momenta, $e$ is the elementary charge and $m$ is the electron mass.

We may derive cross sections for absorption and stimulated emission from the master equation, \cref{eq:MasterEquation}, as follows.
Consider a monoenergetic electron population, such that $f(\vec{p}') = N_e \delta(\vec{p}' - \vec{p})$.
The last term in \cref{eq:MasterEquation} is the absorption contribution:
    \begin{equation}
    \left. \frac{\rmd N(\vec{k})}{\rmd t} \right|_\text{abs} =
        -N_e w(\vec{p}+\vec{k},\vec{k}) N(\vec{k}) =
        -4 \pi^3 n_e \frac{\rmd W(\vec{p}+\vec{k})}{\rmd^3 \vec{k}} N(\vec{k}),
    \label{eq:AbsorptionRate}
    \end{equation}
where we have used $w(\vec{p},\vec{k}) = \frac{(2\pi)^3}{2V} \frac{\rmd W}{\rmd^3 \vec{k}}$ and absorbed the volume factor $V$ into the electron number density, $n_e = N_e / V$.
If the background field is constant and crossed, the triple-differential photon emission rate $\frac{\rmd W}{\rmd^3 \vec{k}}$ is given by~\cite{baier}:
    \begin{equation}
    \frac{\rmd W(\vec{p})}{\rmd^3 \vec{k}} =
        \frac{\alpha}{\sqrt{3} \pi^2 m^2}
        \frac{\zeta^{1/3} (1 + u)}{\gamma^2 \chi_e u}
        \left\{
            \zeta^{2/3} \left[ 1 + (1+u)^2 \right] - (1+u)
        \right\}
        K_{1/3}\!\left( \frac{2 u \zeta}{3 \chi_e} \right),
    \label{eq:TripleDifferential}
    \end{equation}
where
    \begin{align}
    u &= \frac{\omega}{\gamma m - \omega} = \frac{s}{1 - s},
    &
    \zeta &= [2\gamma^2(1-\beta\cos\theta)]^{3/2} =
        \left( \frac{2}{s} \frac{k.p}{m^2} \right)^{3/2}.
    \end{align}
Here $\omega$ is the photon energy, $\gamma$ is the Lorentz factor of an electron with velocity $\beta$, and $\theta$ is the angle between the electron and photon momentum (assumed to be small in the ultrarelativistic limit $\gamma \gg 1$).
The rate at which photons are absorbed depends on the synchrotron emissivity of an electron with momentum $\vec{p} + \vec{k}$~\cite{ghisellini.mnras.1991}.
In order to obtain an equivalent of \cref{eq:TripleDifferential} for an electron with momentum $\vec{p}' = \vec{p} + \vec{k}$, we make the following substitutions: $u \to s$, $\gamma \to \gamma (1 + s)$, $\chi_e \to \chi_e (1 + s)$.
The parameter $\zeta$ is \emph{not} changed.
This may be seen by calculating $k.p'$ in terms of $k.p$, and applying the conservation of momentum in a crossed field with lightlike wavevector $n$: $p + k = p' + [k.p/n.(k+p)] n$.
Dividing \cref{eq:AbsorptionRate} by another volume factor $V$, to obtain a photon number density $n_\gamma$, gives us the number of absorption events per unit volume, per unit time:
    \begin{equation}
    -\left. \frac{\rmd n_\gamma}{\rmd t} \right|_\text{abs} =
    4\pi^3 n_e n_\gamma \frac{\rmd W(\vec{p}+\vec{k})}{\rmd^3 \vec{k}} =
        \underbrace{\frac{n_e n_\gamma}{s \gamma^2} \frac{k.p}{m^2}}_{\text{invariant flux}}
        \underbrace{\frac{4 \pi^2 \alpha}{k.p} \frac{4 g \bar{z} - z}{s} \Ai(\bar{z}) }_{\text{cross section}},
    \end{equation}
where we identify the final result as the product of the invariant flux $F = n_e n_\gamma \, k.p / (k^0 p^0)$ and a cross section $\sigma$.
The auxiliary variables are $s = \chi_\gamma / \chi_e$, $g = \frac{1}{2} + \frac{s^2}{4(1+s)}$, $z = \{s / [\chi_e (1 + s)]\}^{2/3}$ and $\bar{z} = (2 z/s)(k.p/m^2)$.

A similar logic may be followed to obtain the cross section for stimulated synchrotron emission.
From the second term in \cref{eq:MasterEquation},
    \begin{equation}
    \left. \frac{\rmd N(\vec{k})}{\rmd t} \right|_\text{st}
        = N_e w(\vec{p},\vec{k}) N(\vec{k})
        = 4 \pi^3 n_e \frac{\rmd W(\vec{p})}{\rmd^3 \vec{k}} N(\vec{k}),
    \label{eq:StimulatedRate0}
    \end{equation}
we obtain
    \begin{equation}
    \left. \frac{\rmd n_\gamma}{\rmd t} \right|_\text{st} =
        \underbrace{\frac{n_e n_\gamma}{s \gamma^2} \frac{k.p}{m^2}}_{\text{invariant flux}}
        \underbrace{\frac{4 \pi^2 \alpha}{k.p} \frac{4 g' \bar{z}' - z'}{s} \Ai(\bar{z}') }_{\text{cross section}},
    \label{eq:StimulatedRate}
    \end{equation}
where $s = \chi_\gamma / \chi_e$, $g' = \frac{1}{2} + \frac{s^2}{4(1-s)}$, $z' = \{s / [\chi_e (1 - s)]\}^{2/3}$ and $\bar{z}' = (2 z'/s)(k.p/m^2)$.
Note that the same $\vec{k}$ appears on both sides of \cref{eq:StimulatedRate0} and therefore, when stimulated emission occurs, the emitted and stimulating photons have the same momentum.

The cross sections for absorption and stimulated emission may therefore be written an a unified form, as:
	\begin{equation}
	\sigma =
		\frac{4\pi^2\alpha}{k.p} \frac{z(4 g \bar{z}/z - 1) \Ai(\bar{z})}{s},
	\label{eq:CrossSection}
	\end{equation}
where $s = \chi_\gamma / \chi_e$ and $\bar{z} = (2z/s) (k.p / m^2)$ for both processes.
In the remaining two auxiliary variables, $g = 1/2 + s^2/[4(1 \pm s)]$ and $z = \{s / [\chi_e (1 \pm s)]\}^{2/3}$, choosing the positive (negative) sign yields the cross section for absorption (stimulated emission).
The sign of $s$ in the definitions of $g$ and $z$ is an expression of the conservation of momentum.
An electron may absorb a photon with arbitrarily large energy and therefore $s$ may take any value.
However, stimulated emission can only take place for photons with \emph{less} energy than the electron; thus on kinematic grounds, we have the restriction $s < 1$.
The cross section for absorption, obtained in this way, agrees with the result of a direct calculation from strong-field QED~\cite{ilderton.prd.2019}, which is a useful crosscheck of the master equation approach.
To the best of our knowledge, a QED cross section for stimulated emission has not previously been reported.

This result is obtained in the locally constant, crossed field approximation (LCFA), under which the rate for a QED process in an arbitrary background field may be replaced with its equivalent in a constant, crossed field~\cite{ritus.jslr.1985}.
The validity of this approximation depends on the normalized field amplitude $a_0 = e E_0 /(m \omega_0)$, where $E_0$ is the electric field strength and $\omega_0$ is the field's frequency of oscillation.
The LCFA holds for the `1 to 2' processes of Compton scattering~\cite{harvey.pra.2015,dinu.prl.2016,dipiazza.pra.2018,blackburn.pop.2018,ilderton.pra.2019} and nonlinear Breit-Wheeler pair creation~\cite{king.pra.2020} if $a_0$ satisfies $a_0 \gg 1$ and $a_0^3 / \chi_{e,\gamma} \gg 1$, as under these conditions the formation length is much smaller than the scale of variation of the background field.
In a pulsed background, however, there are always temporal regions where the local value of $a_0$ is small, and hence the assumptions of the LCFA are automatically violated.
Compton scattering and Breit-Wheeler pair creation `self-regulate' in this situation~\cite{ilderton.pra.2019};
while the fractional error in the rate is large in such regions, the rate itself is small (in fact, vanishing) due to the behaviour of the Airy functions appearing there, and thus the absolute error is small. The question arises as to what extent these statements apply also to induced processes, which depend on additional kinematic variables.

A comparison of the LCFA for one-photon absorption \cref{eq:CrossSection} with the full QED result~\cite{ilderton.prd.2019} in a monochromatic plane-wave background shows good agreement for $s \gtrsim \chi_e / a_0^3$. Absorption is, though, more likely in regions where
$a_0$ is not large.
In very short pulses, these regions can contribute a significant proportion
of the total probability~\cite{ilderton.prd.2019}.
However, note that \cite{ilderton.prd.2019} benchmarked absorption using
externally injected photons, which overlap with the electrons even
in free space.
Here we consider photons that are emitted by the electron population itself,
so that overlap takes place only in the high-field region, $a_0 \gg 1$, where
emission is most likely.
As the LCFA is satisfied for the emission process in this regime, and emission and absorption take place in the same region of space, it should also be
satisfied for the absorption process.

Emission of a photon by an electron, followed by absorption of that photon by another electron, may be viewed as the component of M{\o}ller scattering ($ee \to ee$, in a strong field) in which the intermediate photon is real.
A complete treatment of electron-electron scattering in a background field would include off-shell and interference contributions;
this has been done for monochromatic~\cite{oleinik.jetp.1967,bergou.jpa.1981,roshchupkin.lp.1996,panek.pra.2004} and pulsed electromagnetic waves~\cite{nedoreshta.pra.2015} at low intensity $a_0 \lesssim 1$, with particular focus on resonances in the transition amplitude.
These resonances occur when the intermediate photon goes on shell, which significantly enhances the interaction probability over its value in vacuum.
This is precisely the interaction under consideration here.
It should dominate the virtual component, i.e. direct electron-electron scattering, which is, in its usual classical description~\cite{melrose.jpp.2007}, negligible for laser-plasmas.

\section{In a laser-plasma environment}

\subsection{Analytical estimates}
\label{sec:Estimates}

Let us first determine the laser and plasma parameters for which one-photon
absorption becomes important.
Consider a population of electrons, with number density $n_e$,
performing a circular orbit with Lorentz factor $\gamma$,
quantum parameter $\chi_e$ and gyroradius $R_c = \gamma^2 / (m \chi_e)$.
Let the space be filled by photons with number density $n_\gamma$,
quantum parameter $\chi_\gamma$ and energy $\omega$,
all propagating in the same direction and in the plane of the electron orbit.

Defining $\theta$ to be the angle between the electron and photon momenta
and assuming $\gamma \gg 1$ and $\theta \ll 1$, the argument of the Airy function in \cref{eq:CrossSection} may be cast as $\bar{z} \simeq \theta^2/\theta_c^2$, for $\theta_c = [m \chi_e / (\gamma^2 \omega)]^{1/3}$.
This shows that the cross section is suppressed for $\theta > \theta_c$, i.e. unless
the electron and photon are almost collinear, so it occurs
once per orbit.
In general, both absorption and stimulated emission are likeliest for low-energy photons propagating at small angles to the electron trajectory.

The number of events per unit volume $n_\text{abs} = \int F \sigma(t) \rmd t$,
where $F = n_e n_\gamma\, k.p / (k^0 p^0)$ is the invariant flux,
$\sigma(t)$ the instantaneous cross section,
and the integral is taken over the interval where $\vec{p}$ is close to parallel
with $\vec{k}$.
Assuming that $s = \chi_\gamma / \chi_e \ll 1$ and the angle between
electron and photon $\theta(t) = t / R_c \ll 1$, we obtain
    \begin{equation}
    F \sigma(t) =
		\frac{4 \pi^2 \alpha n_e n_\gamma}{m^2}
		\frac{1 + 2 \gamma^2 \theta^2}{\gamma^2 \chi_e^{2/3} s^{4/3}}
		\Ai\!\left[
			(s/\chi_e)^{2/3} (1 + \gamma^2 \theta^2)
		\right].
	\label{eq:Integrand}
	\end{equation}
We integrate \cref{eq:Integrand} using the fact that
    \begin{equation}
    \int_{-\infty}^{\infty} (1+2\tau^2) \Ai[\xi (1 + \tau^2)] \,\rmd\tau
        \simeq 0.530 \xi^{-3/2}
    \end{equation}
for $\xi \ll 1$. The fraction of photons absorbed by
the electrons is given by:
    \begin{equation}
    f_\text{abs} = \frac{n_\text{abs}}{n_\gamma} \simeq
        \frac{0.15 n_e}{m^3 \gamma \chi_e^{2/3} s^{7/3}}.
	\label{eq:AbsorptionFraction}
    \end{equation}

In the case that the electrons are in a plasma that is driven
by a circularly polarized laser with angular frequency $\omega_0$,
we can set $\chi_e = \gamma^2 \omega_0/m$
and express the density $n_e$ in terms of the critical density
$\ncrit = m \omega_0^2 / (4 \pi \alpha)$.
We define the self-absorption frequency $\omega_\text{abs}$ as the
largest frequency for which the absorption fraction $f_\text{abs} \gtrsim 1$:
	\begin{equation}
    \omega_\text{abs} [\text{keV}] \simeq
        0.4
        \left(\frac{n_e}{n_\text{cr}}\right)^{3/7}
        \lambda^{-4/7}[\micron]
	\label{eq:ThresholdFrequency}
	\end{equation}
Photons with energies smaller than $\omega_\text{abs}$,
which lies in the multi-keV range for overdense plasmas,
should be efficiently absorbed.

The probability that scattering, via the linear Compton process $e\gamma \to e\gamma$, occurs instead of absorption is negligible for photons satisfying \cref{eq:ThresholdFrequency}.
The fraction of photons scattered $f_\text{sc} = \int F \sigma_\text{sc}\rmd t$, where $F = n_e n_\gamma k.p / (k^0 p^0) \simeq n_e n_\gamma (1 - \cos\theta)$ is the invariant flux and we take $\sigma_\text{sc} = 8\pi\alpha^2/(3m^2)$ as a representative value of the Compton cross section.
The integral over a single orbit yields $f_\text{sc} / f_\text{abs} \simeq 0.019 s^{7/3} \gamma^3 / \chi_e^{1/3}$:
for electrons in a plasma driven by a circularly polarized laser, this is equivalent to $f_\text{sc} / f_\text{abs} \simeq 6.7\times 10^{-7}\omega^{7/3} [\text{keV}] \lambda^{1/3} [\micron]$.
Under these circumstances, it is safe to neglect scattering for the multi-keV photons that are of interest here.

Note that there is no dependence on laser intensity in \cref{eq:ThresholdFrequency}.
The laser intensity does, however, play a role, in that
the origin of the photons that are to be absorbed is
electron synchrotron radiation,
which only becomes substantial if the laser intensity is
sufficiently high~\cite{ridgers.prl.2012,li.pop.2014,nerush.ppcf.2015}.
We now estimate the properties of this emission for the
scenario of a laser-irradiated, overdense plasma.
Only electrons within the skin layer are exposed to strong
electromagnetic fields;
the effective value of the laser amplitude is reduced by
screening from $a_0$, its value in vacuum,
to $a_\text{eff} \simeq a_0 \sqrt{\ncrit/n_e}$~\cite{ridgers.prl.2012}.
(This result strictly applies only in the nonrelativistic limit~\cite{gamaliy.pra.1990},
but it is consistent with the simulation results to be presented.)

Electrons are accelerated on segments of circular trajectories, with Lorentz factor $\gamma \simeq a_\text{eff}$, and emit synchrotron radiation with a characteristic frequency of $\omega_\text{cr} \simeq \gamma^3 \omega_0$.
We expect the LCFA to be valid for the emission and absorption of photons that satisfy $s > \chi_e / a_0^3$, which is equivalent to $\omega > \omega_0$.
This is satisfied for both the self-absorption frequency $\omega_\text{abs}$ and the characteristic frequency of emission $\omega_\text{cr}$:
with $n_e = 100 \ncrit$ and $a_0 = 400$, for example, $\gamma \simeq a_\text{eff} \simeq 40$, $\chi_e = \gamma^2 \omega_0 / m \simeq 5\times 10^{-3}$ and $\omega_\text{cr} \simeq 100$~keV.
The treatment of synchrotron radiation as incoherent requires that the frequencies of interest $\omega > \omega_\text{coh}$, where $\omega_\text{coh} = n_e^{1/3}$ is an upper limit for the onset of coherence effects~\cite{gonoskov.pre.2015}.
Both $\omega_\text{abs}$ and $\omega_\text{cr}$ meet this requirement by at least a factor of two.

The cross sections for stimulated emission and absorption are similar in magnitude for $s \ll 1$~\cite{ghisellini.mnras.1991}.
The balance between the two is determined by the gradient in momentum space of the electron distribution function: net absorption occurs when this is negative, i.e. there are more electrons at lower energy than at higher energy~\cite{melrose}.
This dependence on the electron distribution function means that we turn to numerical methods, i.e. particle-in-cell simulations.

\subsection{Implementation in numerical simulations}
\label{sec:Simulations}

Particle-in-cell simulations now incorporate both the quantum emission and absorption of synchrotron radiation, in addition to classical, relativistic plasma dynamics.
In this work, emission is modelled in the usual Monte Carlo approach~\cite{ridgers.jcp.2014,gonoskov.pre.2015} by integrating the LCFA rate~\cite{erber.rmp.1966,ritus.jslr.1985} along the electron trajectory and sampling the quantum synchrotron spectrum.
We use a spectrum that is differential in both energy and scattering angle~\cite{baier,blackburn.pra.2020}.
Absorption and stimulated emission are incorporated as a binary interaction between macroparticles.

The probabilities of absorption and stimulated emission for an individual macrophoton (index $i$) are controlled by two optical depths $\tau_i^\ell$ ($\ell = \text{abs},\text{stim}$).
To ensure correct statistics, these are initialized with exponentially distributed values, i.e. $\tau_i^\ell = -\ln U_i^\ell$, where the $U_i^\ell$ are pseudorandom numbers chosen on the unit interval~\cite{ridgers.jcp.2014}.
At every timestep, the interaction probability $P^\ell_{ij}$ is calculated for all pairwise combinations of macroelectrons $j$ and macrophotons $i$ that are located in the same grid cell, using the cross sections given in \cref{eq:CrossSection}: $P^\ell_{ij} = w_j (c \Delta t/V) (k_i.p_j/k^0_i p^0_j) \sigma^\ell$, where $w_j$ is the macroelectron weight, $\Delta t$ is the timestep, $V$ is the volume of a grid cell, $k$ is the four-momentum of the photon, and $p$ is the four-momentum of the electron.
Each interaction therefore corresponds to a single absorption (emission) event, rather than to a cumulative treatment of scattering often used for Coulomb collisions~\cite{nanbu.jcp.1998,sentoku.jcp.2008}; a similar scheme has been used to implement linear Compton scattering in PIC simulations~\cite{delgaudio.jpp.2020}.

While the cross sections \cref{eq:CrossSection} are derived for a plane electromagnetic wave in the constant field limit, it is applied to arbitrary background fields in our code.
To do so, we replace $s \to k^0 / p^0$ in the factor of $\bar{z}/z$ appearing in the prefactor. 
(Elsewhere it remains $s = \chi_\gamma / \chi_e$.)
The purpose of this change is to guarantee that the cross section is positive.
We have verified that it does not change the final results of our simulations, as \cref{eq:CrossSection} is strongly suppressed unless the electron and photon are almost collinear.

The macrophoton's optical depths are updated as $\tau^\ell_i \to \tau^\ell_i - P^\ell_{ij}$ for each electron (index $j$), until one of $\tau^\ell_i$ falls below zero.
The relevant interaction is then deemed to occur; in combination with the initialization of $\tau_i^\ell$ described above, this ensures the correct distribution of scattering events~\cite{ridgers.jcp.2014}.
If absorption occurs ($\tau^\text{abs}_i < 0$), the macroelectron momentum is updated as $\vec{p}_j \to \vec{p}_j + w_i \vec{k}_i / w_j$, where $w_i$ is the weight of the macrophoton, and the macrophoton is deleted from the simulation.
(The weight factors appear in order to ensure conservation of momentum.)
If stimulated emission occurs ($\tau^\text{stim}_i < 0$), the macroelectron momentum is updated as $\vec{p}_j \to \vec{p}_j - \vec{k}_i$, a new macrophoton with momentum $\vec{k}_i$ and weight $w_j$ is added to the simulation, and the optical depth of the stimulating photon $\tau^\text{stim}_j$ is reinitialized.
Should both optical depths fall below zero simultaneously, a pseudorandom number $U'$ is drawn on the unit interval and absorption selected if $U' < P^\text{abs}_{ij} / (P^\text{abs}_{ij} + P^\text{stim}_{ij})$; otherwise stimulated emission is selected.
Benchmarking against analytical results are given in \cref{sec:Benchmarking}.

\subsection{Results}
\label{sec:Results}

As an example, we simulate the interaction of a 10-fs (fwhm duration),
circularly polarized laser pulse with a slab of fully ionized carbon plasma,
density $n_e = 100 \ncrit$ and thickness $5.0~\micron$,
at normal incidence.
The laser amplitude is $a_0 = 400$ and its wavelength $\lambda = 800$~nm,
which yields an electron density of $1.7 \times 10^{23}~\text{cm}^{-3}$.
The simulation is performed in 1D, with 1000 cells per micron
and 200 particles per cell for each species.
This neglects the possibility that electrons and photons escape the laser focal spot in one of the transverse directions, as these are ignored in 1D.
However, we show that photons are absorbed (or stimulate emission) in a sufficiently short timescale after emission that the perpendicular distance travelled is small.
Simulations include radiation emission (both spontaneous and stimulated) and absorption by the method discussed in \cref{sec:Simulations}.
The validity of the LCFA in this scenario, upon which this method depends, is discussed in \cref{sec:FormationLength}.
Binary interactions between electrons or photons and ions, such as bremsstrahlung, are suppressed as carbon has a relatively small atomic number $Z = 6$;
as discussed in \cref{sec:Estimates}, linear Compton scattering is negligible.

	\begin{figure}
	\includegraphics[width=0.9\linewidth]{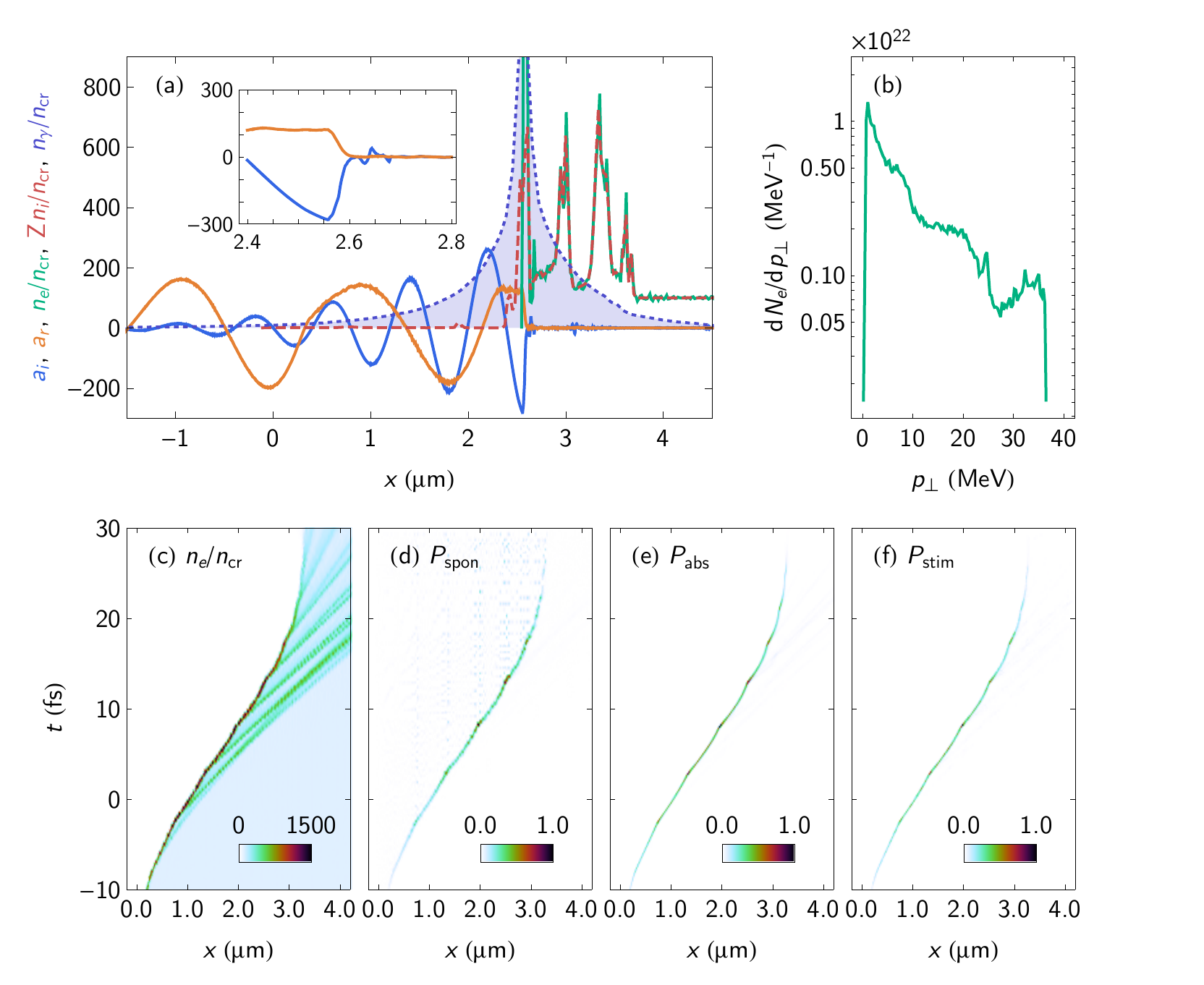}
	\caption{
		(a) The normalized incident and reflected electromagnetic fields (blue and orange) and electron (green), ion (red, dashed) and photon (purple, dashed) number densities at $t = 13.3$~fs.
		(b) Perpendicular momentum distribution of electrons located at positions $x \leq 2.6~\micron$, i.e. within the skin layer, at $t = 13.3$~fs.
		(c) The electron number density as a function of time $t$ and longitudinal coordinate $x$.
		(d) The probability density that a photon is emitted (spontaneously) at time $t$ and position $x$.
		(e) The probability density that a photon, if absorbed, is absorbed at time $t$ and position $x$.
		(f) The probability density that photon emission is stimulated at time $t$ and position $x$.
		In (d-f) all probability densities are normalized to their maxima.}
	\label{fig:PlasmaDynamics}
	\end{figure}

The $y$ components of the incident and reflected electromagnetic field,
as well as the electron, ion and photon number densities at $t = 13.3$~fs
are shown in \cref{fig:PlasmaDynamics}(a).
(Time $t = 0$ corresponds to the centre of the laser pulse
crossing $x = 0$, the location of the unperturbed vacuum interface.)
Electrons near the plasma surface are accelerated on
circular orbits by the laser fields, with perpendicular momenta $p_\perp \simeq m a_\text{eff}$, as shown in \cref{fig:PlasmaDynamics}(b), and displaced by
the radiation pressure in the $x$-direction, as shown in \cref{fig:PlasmaDynamics}(c).
This establishes a charge-separation field that accelerates the ions
in turn.
In the steady state, the velocity of the hole-boring
front is $\beta_\text{hb} = \sqrt{\Xi} / (1 + \sqrt{\Xi})$, where
$\Xi = Z \ncrit m a_0^2 / (A n_e m_p)$~\cite{robinson.ppcf.2009},
$Z$ and $A$ are the atomic and mass numbers of the ion species,
and $m_p$ is the proton mass.
For the parameters under consideration here, $\beta_\text{hb} \simeq 0.40$,
which is consistent with simulation results.

In order to identify when and where induced processes occur, we use the simulation data to calculate the probability density $p(t,x)$ that a photon, if it undergoes absorption or stimulated emission, does so at time $t$ and coordinate $x$.
These two probability densities, along with the equivalent for spontaneous synchrotron emission, are shown in \cref{fig:PlasmaDynamics}(d-f): as all three have unit integral, $\iint\! p(t,x) \,\rmd t \rmd x = 1$, they are scaled by their respective maxima.
\Cref{fig:PlasmaDynamics}(d) shows that synchrotron radiation originates
from electrons in the skin layer, close to the hole-boring front,
where the laser fields are only partially screened.
The skin layer is also where photon absorption and stimulated emission take place [see \cref{fig:PlasmaDynamics}(e) and (f)],
because the photons and electrons are only aligned within an angle of $1/\gamma$
shortly after emission, the local densities are high,
and screening of the background field is not complete.
By recording the time of emission for each photon, as well the momentum, we may calculate the total distance travelled before absorption occurs (including in the transverse directions).
Of all the photons that are absorbed,
90\% are absorbed before they have propagated a distance of $10$~nm.
The smallness of this distance, as compared to the typical size of a laser focal spot, indicates multidimensional effects can safely be neglected.
If the radiation escapes the skin layer, it is highly unlikely to be
absorbed thereafter.

	\begin{figure}
	\includegraphics[width=\linewidth]{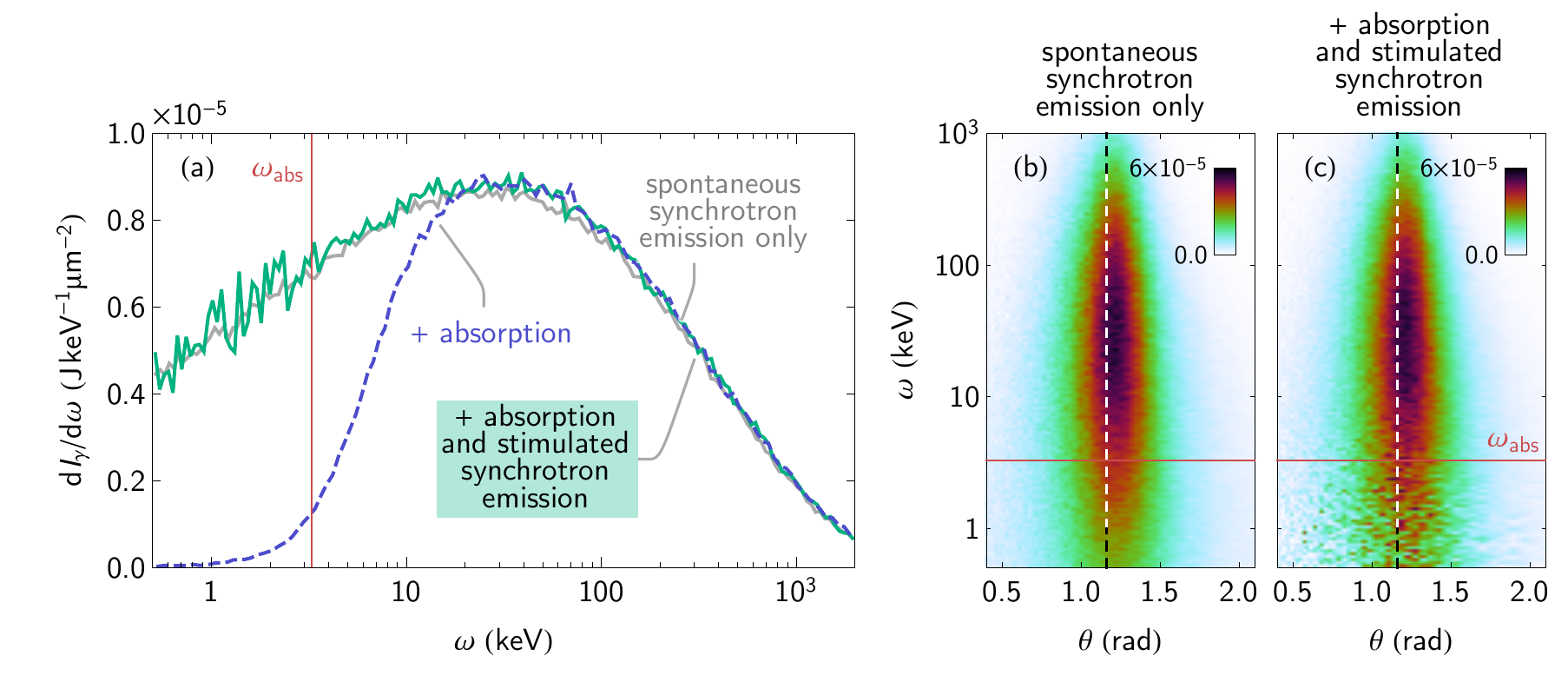}
	\caption{
	    Spectra of the synchrotron photons emitted
	    when plasma with density $n_e = 100 \ncrit$ is irradiated by a
	    circularly polarized laser with peak amplitude $a_0 = 400$:
	    (a) energy radiated per unit frequency, per unit area illuminated,
	    at polar angles $65\degree < \theta < 75\degree$ to the laser axis;
	    (b, c) as in (a), but differential in the polar angle, rather than integrated over it.
	    Solid red lines give $\omega_\text{abs}$, \cref{eq:ThresholdFrequency},
	    our theoretical prediction for the onset of absorption.
	    Dashed lines give $\cos\theta \simeq \beta_\text{hb}$, the expected
	    emission angle from a surface moving at the hole-boring velocity $\beta_\text{hb}$.}
	\label{fig:SelfAbsorption}
	\end{figure}

The radiation spectrum at the end of the simulation, when the plasma is no longer driven by the laser, is shown in \cref{fig:SelfAbsorption}.
As emission takes place when the electron momentum is instantaneously perpendicular to the laser fields, in the rest frame of the plasma surface, we expect the synchrotron radiation to appear predominantly at polar angles $\theta$ satisfying $\cos\theta \simeq \beta_\text{hb}$, where $\beta_\text{hb}$ is the hole-boring velocity.
This is confirmed by \cref{fig:SelfAbsorption}(b) and (c), which show the radiation spectrum as a function of energy and polar angle.
($\theta = 0$ corresponds to forward emission, i.e. parallel to the laser wavevector.)

There is a significant reduction in the number of multi-keV photons when one-photon absorption is taken into account.
The threshold energy at which the spectrum is suppressed is consistent with our theoretical estimate \cref{eq:ThresholdFrequency}, substituting $n_e/\ncrit = 100$.
However, this suppression is countered by stimulated emission, leading to a photon spectrum that is almost identical to the `spontaneous emission only' result.
This also applies to the angularly resolved spectra, shown in \cref{fig:SelfAbsorption}(b) and (c), with the caveat that there is increased statistical noise in the latter.
This arises because, when both absorption and stimulated emission are included, the photon distribution function is effectively resampled at every timestep.
In astrophysical scenarios, it is expected that net absorption causes the spectrum to be suppressed as $\omega^{5/2}$ at $\omega \ll \omega_\text{cr}$~\cite{piran.rmp.2005}, assuming that the electron population has a power-law distribution of energies $\rmd N_e / \rmd \gamma \propto \gamma^{-p}$ ($p > 0$) and that each electron emits and absorbs at a single frequency $\omega_\text{cr}(\gamma)$.
This is not observed here, as the electron perpendicular momentum distribution shown in \cref{fig:PlasmaDynamics}(b), while having negative gradient, is not sufficiently broad.
We expect that similar results would be obtained for a linearly polarized laser, albeit that there would be significantly more synchrotron radiation in the MeV energy range due to increased electron heating.
Absorption and stimulated emission of photons will still occur at the hole-boring front, where the particle density is high and where the electrons and photons are momentarily aligned.

\section{Conclusions}

In this paper we have considered the interplay between the standard strong-field QED process of nonlinear Compton scattering, or spontaneous photon emission, and the particle-particle processes of absorption and stimulated emission.
By constructing cross sections for these processes within the same scheme (based on the locally constant field approximation) used for spontaneous emission, we have shown that it is feasible to include induced, particle-particle processes in simulations of laser-plasma interactions.
This allows us to capture phenomena that are primarily dependent on density.
While photon absorption occurs prolifically for multi-keV synchrotron photons in a laser-plasma interaction, \emph{net} absorption is weak because of stimulated emission.
Our results motivate investigation into the density dependence of QED phenomena in strong fields,
which adds a new axis to the standard parameter space of intensity ($a_0$) and energy ($\chi_{e,\gamma}$).

\begin{acknowledgments}
We are very grateful to John Kirk for helpful discussions during the preparation of this work.
We acknowledge funding from
the Swedish Research Council (grant 2016-03329, M.M.)
and the Engineering and Physical Sciences Research Council (grant EP/S010319/1, A.I., B.K., A.J.M., S.T.).
Simulations were performed on resources provided by the Swedish National Infrastructure for Computing at the High Performance Computing Centre North and National Supercomputer Centre.
\end{acknowledgments}

\section*{Data availability}
The source code for the PIC simulations is available at Ref.~\cite{opal}.
Version 1.5.1, used in this work, and the data necessary to reproduce the simulation results are openly available at Ref.~\cite{dataset}.

\appendix

\section{Benchmarking}
\label{sec:Benchmarking}

    \begin{figure*}
    \centering
    \includegraphics[width=\linewidth]{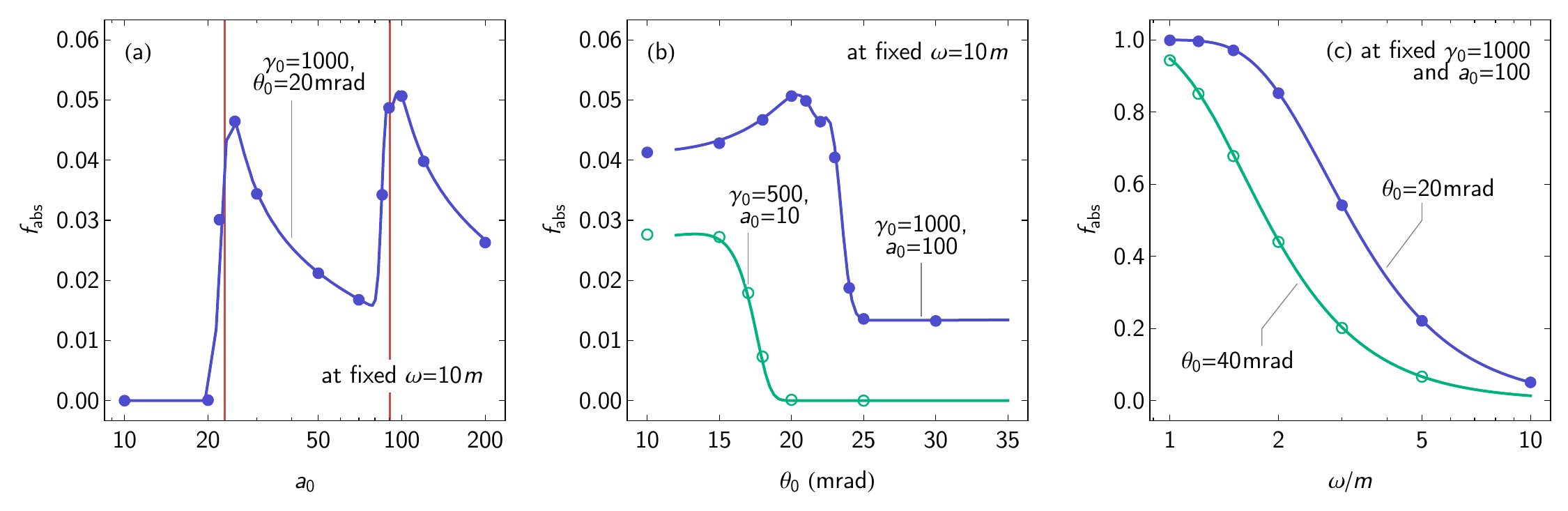}
    \caption{
        Benchmarking of the PIC implementation (points) against analytical predictions (solid lines) for the fraction of photons absorbed $f_\text{abs}$, as a function of (a) laser amplitude $a_0$, (b) collision angle $\theta_0$ and (c) photon energy $\omega$.
        The red vertical lines in (a) indicate the matching conditions
        $\gamma_0 \theta_0 / a_0 = 0.211$ and $0.870$, where
        the electron and photon beams are parallel at
        the field maxima of the laser pulse.
        The initial electron and photon densities are
        $n_e = n_\gamma = 10^{34}~\text{m}^{-3}$.}
    \label{fig:Comparison}
    \end{figure*}

To ensure that the PIC implementation of one-photon absorption outlined in \cref{sec:Simulations} is accurate, we benchmark against the analytical cross section derived in~\cite{ilderton.prd.2019}, for absorption. We consider a linearly polarized plane wave pulse with a $\cos^2$-envelope of duration $\tau \sim 7$~fs, normalized amplitude $a_0$ and wavelength $\lambda_0 = 0.8~\micron$. The potential is given by $e A(\phi) = m a_0 \sin(\phi) \cos^2(\pi \phi/L)$ for phases $\abs{\phi} < L/2$, where $L = 4\pi$. A beam of electrons, with initial energy $\gamma_0 m$ and density $n_e$, and a beam of photons, with energy $\omega$ and density $n_\gamma$, are injected into this pulse: the electron beam counterpropagates into the laser pulse, and we vary the initial angle between the photon beam and laser wavevector $\theta_0$. ($\theta_0 = 0$ corresponds to the electron and photon beams being initially parallel to one another, i.e.~both counterpropagating to the laser.)
Each beam is modelled with 200 macroparticles per cell; we have also verified that varying the number of macroelectrons and macrophotons per cell from (400, 200), to (200, 200), and then to (200, 400), does not alter the results.

A suitable observable is the fraction, $f_{\text{abs}}$, of photons absorbed from the initial beam. Analytically, this is given by
    \begin{equation}
    f_\text{abs} =
        1 - \exp\!\left(-n_e \sigma_\text{int} \tau \frac{1 - \cos\theta_0}{1 + \cos\theta_0}\right),
    \label{eq:exFrac}
    \end{equation}
where $\sigma_\text{int} = \frac{1}{L} \int_{-L/2}^{L/2} \sigma(\phi)\,\rmd\phi$ is the integrated cross section (Eq. 32 in \cite{ilderton.prd.2019}) and $\tau = L/\omega_0$ is the laser duration.
In \cref{fig:Comparison} we compare the fraction of absorbed photons \cref{eq:exFrac} using   $\sigma_\text{int}$ calculated analytically from~\cite{ilderton.prd.2019}, with that obtained numerically by the PIC simulations outlined in \cref{sec:Simulations}. To ensure a fair comparison, photon emission (both spontaneous and stimulated) and current deposition are disabled in the simulations.

The results of our PIC implementation (points) show excellent agreement with the analytical predictions (solid lines) over parameter scans in the field strength $a_0$, initial photon beam angle $\theta_0$, and energy $\omega$. In particular, the PIC implementation correctly resolves the peak structure seen in the dependence of the absorbed fraction $f_{\text{abs}}$ on the field strength $a_0$. These peaks arise when the electrons and photons are brought into alignment at a local maximum of the field amplitude, i.e., when the instantaneous angle between the electron momentum and the laser wavevector, $\theta_e(\phi) \simeq e A(\phi) / (m \gamma_0)$, satisfies $\theta_e(\phi) = \theta_0$, at a phase $\phi$ where $\partial_\phi A(\phi) = 0$. For the two-cycle pulse under consideration here, the matching condition is $\gamma_0 \theta_0 / a_0 = 0.211$ and $0.870$.

The densities employed to generate \cref{fig:Comparison}, $n_e = n_\gamma = 10^{34} \text{m}^{-3}$, are sufficiently high that ignoring current deposition is unphysical. However, as discussed in the main text above, one can alleviate this problem by considering the absorption of synchrotron photons generated in the hole-boring regime. The simulations discussed in the main text do include the fields generated by the plasma.

\section{Finite formation length effects}
\label{sec:FormationLength}

    \begin{figure*}
    \centering
    \includegraphics[width=0.6\linewidth]{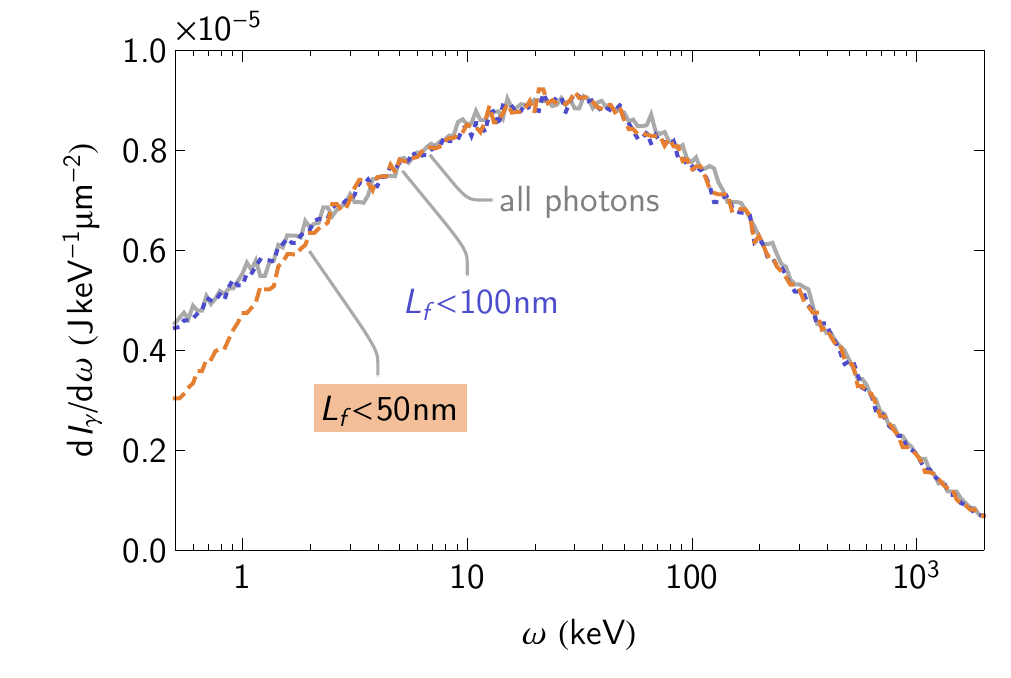}
    \caption{
        Spectra of the synchrotron photons emitted when plasma with density $n_e = 100 \ncrit$ is irradiated by a circularly polarized laser with peak amplitude $a_0 = 400$, as shown in \cref{fig:SelfAbsorption}:
        all photons (grey), photons with formation lengths $L_f$ smaller than 100~nm (purple, dotted) and 50~nm (orange, dashed).}
    \label{fig:FormationLength}
    \end{figure*}

In \cref{sec:Results}, we present PIC simulations of a laser-plasma interaction, which include spontaneous and stimulated synchrotron emission, as well as absorption.
The rates (cross sections) for these processes are calculated within the locally constant field approximation (LCFA).
Physically, this requires that the photon `formation length', the characteristic distance over which emission takes place, be smaller than the spatial scale of variation of the external electromagnetic field~\cite{ritus.jslr.1985}.
The failure of the LCFA at small photon energies or low intensity has motivated a search for photon emission rates that are valid for non-constant backgrounds and appropriate for inclusion in simulations: various methods are now available~\cite{dipiazza.pra.2018,ilderton.pra.2019}.

In this section we estimate the error made by our simulations in using the LCFA, by means of the method described in \cite{blackburn.pra.2020}.
This exploits the fact that the LCFA photon emission rate sampled by the code is differential in both energy and angle.
Consequently, the formation length $L_f$ may be estimated as the distance travelled by the electron before it is deflected by an angle $\vartheta$, where $\vartheta$ is the angle between the photon and electron momenta at the point of emission.
We calculate $L_f$ for each simulated photon, using $L_f \simeq r_c \vartheta$, where $r_c \simeq \gamma^2 / (m \chi_e)$ is the electron's instantaneous radius of curvature.
Photons with formation lengths above a threshold value are then discarded in order to estimate the importance of nonlocal effects.
As interference effects do not, in reality, completely suppress emission, this scheme tends to overestimate the necessary correction~\cite{blackburn.pra.2020}.

What value the maximum formation length $L_f^\text{max}$ should take depends on the scenario in question.
In the present case, we use the fact that the skin depth, the length over which the incident electromagnetic fields decay at the plasma surface, shown in the inset of \cref{fig:PlasmaDynamics}(a), is approximately $\lambda_p \simeq 100$~nm.
Photons are emitted primarily at large polar angles $\cos\theta \simeq \beta_\text{hb}$ to the laser axis, where $\beta_\text{hb}$ is the hole-boring velocity: thus an appropriate threshold would be $L_f^\text{max} \simeq \lambda_p / \beta_\text{hb} \simeq 240$~nm (recall $\beta_\text{hb} \simeq 0.40$ for a carbon plasma with density $n_e = 100 \ncrit$).
\Cref{fig:FormationLength} shows the photon spectra (spontaneous, i.e. synchrotron emission only) from the simulations for three different values of $L_f^\text{max}$.
We see that excluding all photons with formation lengths greater than 100~nm, or even 50~nm, has a small overall effect on the spectrum in the few-keV range.
As photons must be emitted in order to be absorbed, or to stimulate further synchrotron emission, we conclude that the LCFA is a reasonable approximation for the interaction studied in this work.

\bibliography{references}

\end{document}